\documentclass{emulateapj}

\usepackage{amsmath,natbib,graphicx}
\usepackage{epsf}
\usepackage{fancyvrb}
\usepackage{epstopdf}
\usepackage{epsfig}
\usepackage{color}
\DeclareGraphicsExtensions{.jpg,.pdf,.png,.eps,.ps}

\begin{document}
\VerbatimFootnotes

\title{Indication of insensitivity of planetary weathering behavior
  and habitable zone to surface land fraction} \shorttitle{Land
  Fraction and Habitability}


\author{Dorian~S.~Abbot\altaffilmark{1}, Nicolas~B. Cowan\altaffilmark{2}, and Fred~J.~Ciesla\altaffilmark{1}}

\altaffiltext{1}{Department of the Geophysical Sciences, University of
  Chicago, 5734 South Ellis Avenue, Chicago, IL 60637, USA;
  abbot@uchicago.edu} \altaffiltext{2}{Center for Interdisciplinary
  Exploration and Research in Astrophysics (CIERA) and Department of
  Physics \& Astronomy, Northwestern University, 2131 Tech Drive,
  Evanston, IL 60208, USA}

\shortauthors{Abbot, Cowan, \& Ciesla}

\email{abbot@uchicago.edu}


\begin{abstract}
  It is likely that unambiguous habitable zone terrestrial planets of
  unknown water content will soon be discovered. Water content helps
  determine surface land fraction, which influences planetary
  weathering behavior. This is important because the silicate
  weathering feedback determines the width of the habitable zone in
  space and time.  Here a low-order model of weathering and climate,
  useful for gaining qualitative understanding, is developed to
  examine climate evolution for planets of various land-ocean
  fractions.  It is pointed out that, if seafloor weathering does not
  depend directly on surface temperature, there can be no
  weathering-climate feedback on a waterworld.  This would
  dramatically narrow the habitable zone of a waterworld. Results from
  our model indicate that weathering behavior does not depend strongly
  on land fraction for partially ocean-covered planets.  This is
  powerful because it suggests that previous habitable zone theory is
  robust to changes in land fraction, as long as there is some
  land. Finally, a mechanism is proposed for a waterworld to prevent
  complete water loss during a moist greenhouse through rapid
  weathering of exposed continents.  This process is named a
  ``waterworld self-arrest,'' and it implies that waterworlds can go
  through a moist greenhouse stage and end up as planets like Earth
  with partial ocean coverage. This work stresses the importance of
  surface and geologic effects, in addition to the usual incident
  stellar flux, for habitability.
\end{abstract}

\bigskip\bigskip

\section{Introduction}\label{sec:intro}

The habitable zone is traditionally defined as the region around a
star where liquid water can exist at the surface of a planet
\citep{Kasting93}. Since climate systems include both positive and
negative feedbacks, a planet's surface temperature is non-trivially
related to incident stellar flux.  The inner edge of the habitable
zone is defined by the ``moist greenhouse,'' which occurs if a planet
becomes hot enough (surface temperature of $\approx$340~K) that large
amounts of water can be lost through disassociation by photolysis in
the stratosphere and hydrogen escape to space. The outer edge occurs
when CO$_2$ reaches a high enough pressure that it can no longer
provide warming, either because of increased Rayleigh scattering or
because it condenses at the surface, which results in permanent global
glaciation.  These limits do not necessarily represent hard limits on
life of all types. For example, life survived glaciations that may
have been global (``Snowball Earth'') on Earth 600--700 million years
ago \citep{Kirschvink92,Hoffman98}. Furthermore, if greenhouse gases
other than CO$_2$ are considered, e.g., hydrogen
\citep{Pierrehumbert:2011p3366,Wordsworth2012-transient}, the outer
limit of the habitable zone could be pushed beyond the CO$_2$
condensation limit. Alternative limits on habitability have also been
proposed. For example, a reduction in the partial pressure of
atmospheric CO$_2$ below $10^{-5}$ bar may prevent C$_4$
photosynthesis, which could curb the complex biosphere
\citep{CALDEIRA:1992p2366,VonBloh:2005p2325}, although it would
certainly not limit most types of life.

Studies of planetary habitability have become increasingly pertinent
as capabilities for the discovery and characterization of terrestrial
or possibly terrestrial exoplanets in or near the habitable zone have
improved. For example, planets GJ581d
\citep{Udry:2007p3239,Mayor:2009p3227,Vogt:2010p3247,Wordsworth:2011p3221},
HD85512b \citep{Pepe11,Kaltenegger11}, Kepler-22b \citep{Borucki11},
and GJ 677c \citep{Anglada12} have recently been discovered and
determined to be potentially in the habitable zone.  It is likely that
the number of good candidates for habitable terrestrial exoplanets
will increase in the near future as the Kepler mission, ground-based
transit surveys, radial velocity surveys, and gravitational
microlensing studies continue to detect new planetary systems.

The ``Faint Young Sun Problem'' discussed in the Earth Science
literature is analogous to the habitable zone concept. A planet
orbiting a main-sequence star receives ever increasing incident flux
(``insolation'') as the star ages. This should tend to hurry a planet
through the habitable zone unless the planetary albedo or thermal
optical thickness (greenhouse effect) adjust as the system ages.  As
pointed out by \citet{SAGAN:1972p1233}, geological evidence for liquid
water early in Earth's history suggests that just such an adjustment
must have occurred on Earth. Although there is some debate about the
details \citep{Kasting:2010p1226}, it is fairly widely accepted that
the silicate-weathering feedback
\citep{Walker-Hays-Kasting-1981:negative} played an important role in
maintaining clement conditions through Earth's history
\citep{Feulner:2012p3570}. Through this negative feedback the
temperature-dependent weathering of silicate rocks on continents,
which represents the main removal process for CO$_2$ from the
atmosphere, is reduced when the temperature decreases, creating a
buffering on changes in temperature since CO$_2$ is a strong infrared
absorber. The silicate weathering feedback also greatly expands the
habitable zone annulus around a star in space \citep{Kasting93}, since
moving a planet further from the star decreases the insolation it
receives just as moving backward in time does. If insolation
boundaries are used to demarcate the habitable zone, the concept can
be used to consider habitability as a function of position relative to
the star at a particular time, or as a function of time for a planet
at a constant distance from a star.

In addition to continental silicate weathering, weathering can also
occur in hydrothermal systems in the basaltic oceanic crust at the
seafloor. Seafloor weathering is poorly constrained on Earth; however,
it is thought to be weaker than continental weathering and to depend
mainly on ocean chemistry, pH, and circulation of seawater through
basaltic crust, rather than directly on surface climate
\citep{CALDEIRA:1995p2285,Sleep:2001p3368,Lehir08}. If this is true,
CO$_2$ would be less efficiently removed from the atmosphere of a
planet with a lower land fraction, leading to higher CO$_2$ levels and
a warmer climate.  Furthermore, a planet with a lower land fraction
could have a weaker buffering to changes in insolation than a planet
with a higher land fraction, which would cause it to have a smaller
habitable zone.

While the weathering behavior of a planet could depend on surface land
fraction, the water complement of planets in the habitable zone should
vary substantially. The reason for this is that the habitable zone is
in general located closer to the star than the snow line, the location
within a protoplanetary disk outside of which water ice would be
present and available to be incorporated into solids. In the solar
system, for example, the current habitable zone ranges from
approximately 0.8 to 1.7 AU \citep{Kasting93} if one neglects the
uncertain effects of CO$_2$ clouds. If CO$_2$ clouds are assumed to
produce a strong warming \citep{Forget:1997p3442}, the outer limit of
the habitable zone could be extended to $\approx$2.4 AU
\citep{Mischna:2000p3444,Selsis:2007p3441,Kaltenegger:2011p3443},
although recent three-dimensional simulations including atmospheric
CO$_2$ condensation suggest that the warming effect may in fact be
fairly modest
\citep{Wordsworth:2011p3221,Forget2012,Wordsworth2012}. The snow line
is thought to have been around $\sim$2.5 AU \citep{morbidelli00};
beyond this distance, evidence for water in the form of ice or
hydrated minerals is seen in asteroids and planets.  The general
picture for water delivery to Earth has been that the orbits of bodies
from beyond the snow line were excited to high eccentricities through
gravitational scattering by massive planetary embryos and a young
Jupiter.  These high eccentricities would have put water-bearing
planetesimals and embryos on Earth-crossing orbits.  A fraction of
such bodies would have been accreted by Earth, stochastically
delivering volatiles to the young planet
\citep{morbidelli00,obrien06,Raymond:2009p2536}.

Efforts have been made to extend this analysis to planetary systems
around other stars.  The same dynamical effects are expected to occur,
with the amount of water delivered to a potentially habitable planet
being strongly dependent on the presence or orbits of giant planets
\citep{obrien06,raymond06,Raymond:2009p2536}.  As low mass stars are
expected to have lower mass disks and therefore fewer massive bodies
to produce gravitational scattering, low mass stars may be more likely
to have volatile-poor habitable-zone planets \citep{Raymond:2007}.
Stars of mass $\sim$1~$M_{\odot}$ or higher, on the other hand, could
easily have habitable-zone planets of similar or greater fractional
mass of water than Earth \citep{Raymond:2007}.

As described above, current thinking mostly focuses on hydrated
asteroids as the main source of water for a habitable planet; however,
there are other dynamical mechanisms that could allow habitable
planets to accrete significant amounts of water and volatiles.
Cometary-bodies would deliver a larger amount of water for a given
impactor mass than asteroids.  Should such bodies get scattered over
the course of planet formation, this may allow larger amounts of water
to be accreted than otherwise predicted.  Furthermore,
\citet{kutchner03} argued that planets which formed outside the snow
line could migrate inwards due to gravitational torques from a
protoplanetary disk or scattering with other planets.  Such bodies
would naturally accrete large fractions of water ice during formation
beyond the snow line, which would then melt and sublimate in a warmer
orbit, providing a high volatile content to a planet close to its
star.

Habitable zone planets could therefore have a wide variety of water
mass fractions, which would lead to varying land
fractions. Furthermore, even for a constant mass fraction of water,
scaling relations dictate that land fraction depends on planetary
size. At the same time, the surface land fraction could exert strong
control on the planetary carbon cycle, which strongly influences
planetary habitability. This warrants a general consideration of
weathering behavior on planets of varying land fraction.

The main objective of this paper is to investigate the effect of land
fraction on the carbon cycle and weathering behavior of a terrestrial
planet in the habitable zone. We will outline and use a simple
analytical model for weathering and global climate that necessarily
makes grave approximations to the real physical processes. For
example, we will use existing parameterizations of seafloor
weathering, while acknowledging that observational and experimental
constraints on such parameterizations are minimal. We will only use
the model, however, to make statements that do not depend strongly on
uncertain aspects of the parameterizations. This model should be used
to understand intuitively the qualitative behavior of the system
rather than to make quantitative estimates. A major strength of the
model is that it is easy to derive and understand, yet should capture
the most significant physical processes. This type of modeling is
appropriate in the study of exoplanets, for which limited data that
would be relevant for a geochemical model exist.

We will consider an Earth-like planet with silicate rocks, a large
reservoir of carbon in carbonate rocks, and at least some surface
ocean. We will use equilibrium relations for weathering, which is
reasonable for the slow changes in insolation that a main-sequence
star experiences. The climate model we use is a linearization of a
zero dimensional model. Although this is a severe approximation, it
allows the analytical progress that we feel is useful for obtaining
insight into the problem. We will consider planetary surface land
fractions ranging from partial ocean coverage to complete
waterworlds. In this context a planet would be a waterworld if the
highest land were covered by even 1~m of water, although a planet with
more water than this would qualify as a waterworld as well. However we
will assume the geophysical context is a planet with a substantial
rock mantle and only up to roughly $\approx$10 times Earth's water
mass fraction \citep[0.02\%-0.1\%;][]{Hirschmann:2012p3481} rather
than potential waterworlds that might be \cal{O}(10\%) or more water
by mass \citep{Fu2010} and would therefore have vastly different
volatile cycles. Although partially ocean-covered planets have
recently been referred to as ``aquaplanets'' \citep{abe2011}, we will
not adopt this terminology because ``aquaplanet'' has a long history
of being used to denote a completely ocean-covered planet in the
climate and atmospheric dynamics communities.

We will find that the weathering behavior is fairly insensitive to
land fraction when there is partial ocean coverage. For example, we
will find that weathering feedbacks function similarly, yielding a
habitable zone of similar width, if a planet has a land fraction of
0.3 (like modern Earth) or 0.01 (equivalent to the combined size of
Greenland and Mexico). In contrast, we will find that the weathering
behavior of a waterworld is drastically different from a planet with
partial ocean coverage. If seafloor weathering depends mainly on ocean
acidity, rather than planetary surface temperature, no weathering
feedback operates on a waterworld and it should have a narrow
habitable zone and progress through it quickly as its star
ages. Finally, we will argue that it is possible for a waterworld to
stop a moist greenhouse in progress when continent is exposed by
drawing down the CO$_2$ through massive weathering, which would leave
the planet in a clement state with partial ocean coverage. We will
refer to this possibility as a ``waterworld self-arrest.''

This paper complements in two ways the recent work of \citet{abe2011},
who found that a nearly dry planet should have a wider habitable zone
than a planet with some water. First, the calculations made by
\citet{abe2011} focused on climate modeling rather than weathering,
although they included a qualitative description of factors that would
influence weathering on a dry planet. Second, we consider a variety of
land fractions up to the limiting case of a waterworld.

The outline of this paper is as follows. We describe our model in
Section \ref{sec:model}, use it in Section \ref{sec:results}, and
perform a sensitivity analysis in Section \ref{sec:sensitivity}. We
discuss the possibility of a waterworld self-arrest in Section
\ref{sec:self-arrest}. We outline observational implications and
prospects for confirmation or falsification of our work in Section
\ref{sec:observables}. We discuss our results further in Section
\ref{sec:discussion}, including considering the limitations of our
various assumptions, and conclude in Section \ref{sec:conclusions}.

\section{Model Description}\label{sec:model}

Here we will briefly review silicate weathering and the carbon cycle
before developing our model. The reader interested in more detail
should consult \citet{Berner2004} and Chapter 8 of
\citet{Pierrehumbert:2010-book}. The carbon cycle on a terrestrial
planet can be described as a long-term balance between volcanic
outgassing of CO$_2$ and the burial of carbonates, mediated by
chemical weathering of silicates. Silicate minerals are weathered
through reactions such as
\begin{equation}
\label{eq:weathering_chemistry}
\mathrm{CaSiO_3(\mathrm{s})+CO_2(\mathrm{g})} \rightleftharpoons 
\mathrm{CaCO_3(\mathrm{s}) + SiO_2(\mathrm{s})},
\end{equation}
where we have used CaSiO$_3$ as an example silicate mineral, but
MgSiO$_3$, FeSiO$_3$, and much more complex minerals can also
participate in similar weathering reactions. The reaction product
CaCO$_3$ is an example carbonate and silicon dioxide (SiO$_2$) is
often called silica. These reactions occur in aqueous solution and
lead to a net decrease in atmospheric CO$_2$ if the resulting CaCO$_3$
is eventually buried in ocean sediment, ultimately to be subducted
into the mantle.  CaCO$_3$ is generally produced in the ocean by
biological precipitation, but if there were no biology weathering
fluxes into the ocean would drive up carbonate saturation until
abiotic CaCO$_3$ precipitation were possible.  Carbon in the mantle
can be released from carbonate minerals if it reaches high enough
temperatures, leading to volcanic outgassing and closing the carbon
cycle. The mantle reservoir of carbon is large enough that CO$_2$
outgassing does not depend on the amount of carbon subducted into the
mantle.

Weathering reactions can occur either on continents or at the seafloor
and the weathering rate is the total amount of CO$_2$ per year that is
converted into carbonate by reactions like
Equation~(\ref{eq:weathering_chemistry}) and subducted. The weathering rate
is often measured in units of kilograms carbon per year
(kg~C~yr$^{-1}$). Below we will work with the dimensionless weathering
rate, which is normalized such that it equals unity at the modern
weathering rate. Because weathering reactions are aqueous and because
rain increases erosion, allowing more rock to react, the continental
weathering rate should increase with increasing
precipitation. Experiments suggest that weathering reactions should
increase with increasing CO$_2$ concentration, but that this
dependence may be significantly reduced when land plants are present
\citep{Berner2004,Pierrehumbert:2010-book}. Finally, weathering rates
are generally assumed to have an exponential dependence on
temperature. This is based on temperature fits
\citep{MARSHALL:1988p3303} to cation, e.g., Ca$^{2+}$ and Mg$^{2+}$,
data in river runoff \citep{meybeck1979}, and appears to be ultimately
due to an Arrhenius law for rock dissolution \citep{Berner:1994p3295}.

These dependencies of continental weathering can be combined into the
following approximation for $W_l$, the continental silicate weathering
rate,
\begin{equation}
\label{eq:weathering}
\frac{\tilde{W_l}}{W_{l0}}=\left( \frac{\tilde{P}}{P_0} \right)^a
\left( \frac{\tilde{\phi}}{\phi_0} \right)^b
e^{\frac{\tilde{T}-T_0}{T_u}},
\end{equation}
where $P$ is the precipitation rate, $\phi$ is the partial pressure of
CO$_2$, $T$ is the surface temperature (see Table~\ref{tab:params} for
a list of important model variables), and $T_u=10^\circ$~C, $a=0.65$,
and $b=0.5$ are constants. The subscript $0$ represents the current
value and a tilde represents a dimensionful quantity.  The
precipitation dependence of continental weathering, $a$, is determined
experimentally by measuring the Ca$^{2+}$ and Mg$^{2+}$ solute
concentration in rivers and regressing against annual runoff.
\citet{Berner:1994p3295} and \citet{Pierrehumbert:2010-book} use
$a=0.65$ based on a study in Kenya that found $a=0.6$
\citep{DUNNE:1978p3297} and a study in the United States that found
$a=0.7$ \citep{Peters84}. We will use $a=0.65$ as our standard value,
but we find in Section \ref{sec:sensitivity} that our main results are
robust for $a$ between 0 and 2. The CO$_2$-dependence of continental
weathering, $b$, could be as low as 0 if land plants eliminate the
effect of CO$_2$ on weathering \citep{Pierrehumbert:2010-book}, and
the maximum value it should take is 1 \citep{Berner:1994p3295}. Both
\citep{Berner:1994p3295} and \citet{Pierrehumbert:2010-book} adopt the
intermediate value of 0.5, as we will here. Our main results are only
sensitive to $b$ if it becomes very small (Section
\ref{sec:sensitivity}). Various estimates of the $e$-folding
temperature, $T_u$, exist in the literature, including $T_u=10.0$~K
\citep{Pierrehumbert:2010-book}, $T_u=11.1$~K
\citep{Berner:1994p3295}, $T_u=21.7$~K \citep{MARSHALL:1988p3303}, and
$T_u=17.7$~K \citep{Walker-Hays-Kasting-1981:negative}. We adopt
$T_u=10.0$~K and show that our main conclusions are valid for
$T_u=5-25$~K in Section \ref{sec:sensitivity}.

Equation~(\ref{eq:weathering}) can be nondimensionalized to take the
following form
\begin{equation}
\label{eq:weathering_nondim}
W_l=P^a\phi^be^T,
\end{equation}
where $W_l=\frac{\tilde{W_l}}{W_{l0}}$, $P=\frac{\tilde{P}}{P_0}$,
$\phi=\frac{\tilde{\phi}}{\phi_0}$, and $T=\frac{\tilde{T}-T_0}{T_u}$.

We parameterize the precipitation as a function of temperature as
\begin{equation}
\label{eq:precip_dim}
\frac{\tilde{P}}{P_0}=1+\tilde{\alpha}_p(\tilde{T}-T_0),
\end{equation}
where $\tilde{\alpha}_p$ represents the fractional increase in
precipitation per degree of warming. Climate change simulations in a
variety of global climate models suggest that
$\tilde{\alpha}_p=0.02-0.03$, corresponding to a 2\%-3\% increase in
global mean precipitation per Kelvin increase in global mean
temperature, is an appropriate value when $\tilde{T} \approx T_0$
\citep{Schneider:2010p3251}. Simulations in an idealized global
climate model, however, suggest that precipitation asymptotically
approaches an energetically determined constant limit at
$\tilde{T}\approx 300-310$~K as the atmospheric optical thickness is
increased \citep{OGorman:2008p3252}. The energetic constraint on
precipitation would be expected to change, however, if the insolation
changes, which is the situation we will be considering below.  For
simplicity, we will take $\tilde{\alpha}_p=0.025$ in what
follows. Because of the exponential dependence of continental
weathering on $T$ in Equation~(\ref{eq:weathering_nondim}), our
results are only minimally affected by $\tilde{\alpha}_p$, such that
our results are not substantially altered even if we eliminate the
temperature dependence of precipitation by setting
$\tilde{\alpha}_p=0$ (Section \ref{sec:sensitivity}). Defining
$\alpha_p=\tilde{\alpha}_pT_u=0.25$ we can nondimensionalize
Equation~(\ref{eq:precip_dim}) to find
\begin{equation}
\label{eq:precip}
P=1+\alpha_pT.
\end{equation}

\begin{table}
  \caption{List of the Most Important Model Variables}
\label{tab:params}
\centering
\begin{tabular}{ll}
  Variables & Definition \\
  \hline 
  $T$ & Global mean surface temperature \\
  $\phi$ & Partial pressure CO$_2$ in the atmosphere \\
  $S$ & Stellar flux at planetary distance \\
  & (solar constant for Earth) \\
  $\gamma$ & Planetary land fraction \\
  $\frac{ \partial T}{\partial S}$ & Climate sensitivity to external forcing \\
  $W$ & Weathering rate \\
  $\Gamma$ & Outgassing rate 
\end{tabular}
\end{table}

In addition to continental silicate weathering, reactions similar to
Equation~(\ref{eq:weathering_chemistry}), which lead to the net consumption
of CO$_2$, can occur at the seafloor. We will refer to this
``low-temperature'' alteration of basaltic ocean crust as ``seafloor
weathering.'' We do not consider ``high-temperature'' basaltic
alteration here because it does not lead to net CO$_2$
consumption. Seafloor weathering should depend on many factors,
including ocean pH, the chemical composition of the ocean, the
temperature at which the reactions occur, circulation of seawater
through basaltic crust, and the seafloor spreading rate.  We will
consider a constant seafloor spreading rate and assume a chemical
composition similar to the modern ocean. Based on laboratory
measurements, \citet{CALDEIRA:1995p2285} parameterized the seafloor
weathering rate ($W_o$) as
\begin{equation}
\label{eq:caldeira}
\frac{\tilde{W}_o}{W_{o0}}= \frac{[\tilde{\mathrm{H}}^+]^m+\tilde{\nu}}{[\mathrm{H}^+]_0^m+\tilde{\nu}},
\end{equation}
where $[\mathrm{H}^+]$ is the hydrogen ion concentration and $m$ is a
constant.  $\tilde{\nu}$ has a value of $10^{-3}$ and units of
M$^{m}$, where M is molar concentration, in order to satisfy
dimensional constraints in Equation~(\ref{eq:caldeira}).  This
parameterization assumes that there is enough weatherable material at
the seafloor, and sufficient circulation of seawater through it, to
accommodate arbitrary increases in ocean acidity. If the weatherable
material produced at ocean spreading regions is completely used up
before subduction occurs, $W_o$ will be limited in a way not included
in Equation~(\ref{eq:caldeira}). We will return to this assumption in
Section \ref{sec:discussion}.

Ocean chemistry leads to a functional relationship between $\phi$, the
atmospheric CO$_2$ concentration, and [H$^+$] of the ocean. We will
assume that the ocean chemistry can be described as a carbonate ion
system coupled to an equilibrium relation for CaCO$_3$. Although this
assumption neglects other sources of ocean alkalinity, e.g.,
Mg$^{2+}$, Na$^{+}$, and K$^{+}$, it has been useful for modeling the
carbon cycle in deep time Earth problems
\citep[e.g.,][]{Higgins03}. Furthermore, we implicitly assume that
Ca$^{2+}$ can be supplied to the ocean by submarine rocks in a
waterworld or a planet with very low land fraction. This system is
described by the equilibrium relations for the dissolution of CO$_2$
and CaCO$_3$, the disassociation of H$_2$CO$_3$, HCO$_3^-$, and
H$_2$O, and alkalinity (charge) balance \citep[see chapter 4 of][for
details]{Drever88}. Using the equilibrium constants given by
\citet{Drever88}, we find that for an atmospheric CO$_2$ concentration
between $10^{-4}$ and $10^{2}$ bar, the dominant alkalinity balance is
$2[Ca^{2+}]=[HCO_3^-]$ in this system. Since in the carbonate ion
system $[Ca^{2+}] \propto \frac{[H^+]}{[HCO_3^-]}$ and $[HCO_3^-]
\propto \frac{\tilde{\phi}}{[H^+]}$, this dominant balance leads to
$[\mathrm{H}^+] = c(T) \tilde{\phi}^\frac{2}{3}$ with $c(T)$ given by a
combination of equilibrium constants. $c(T)$ varies by only
$\cal{O}$(10\%) between 0$^\circ$C and 90$^\circ$C, and we therefore
neglect this temperature dependence below. This leads to the
nondimensional approximation $[\mathrm{H}^+]=\phi^\frac{2}{3}$, which allows
Equation~(\ref{eq:caldeira}) to be written in nondimensional form
\begin{equation}
\label{eq:sf_weat2}
W_o=\left( \frac{\phi^n+\nu}{1+\nu} \right),
\end{equation}
where $W_o=\frac{\tilde{W}_o}{W_{o0}}$, $n=\frac{2}{3}m$, and
$\nu=\frac{\tilde{\nu}}{[H^+]_0^m}$.  Using the
\citet{CALDEIRA:1995p2285} values ($m=\frac{1}{2}$ and
$R=10^{-3}$~M$^{\frac{1}{2}}$) and assuming a reference ocean pH
(pH=-log$[\tilde{\mathrm{H}}^+]$) of 8.2 yields $n=\frac{1}{3}$ and
$\nu$=12.6. Note that \citet{CALDEIRA:1995p2285} discussed the
possibility that an exponent to $[H^+]$ other than $\frac{1}{2}$ may
be appropriate and we will discuss other scalings in Sections
\ref{sec:sensitivity} and \ref{sec:discussion}. We note that we have
neglected any dependence of seafloor weathering on surface
temperature, which is equivalent \citep[e.g.,][]{Sleep:2001p3368} to
assuming the temperature at which seafloor weathering occurs
($\approx$20-40$^\circ$C) is set by heat flux from Earth's interior,
rather than by the temperature of seawater. Since this assumption
eliminates the potential for a weathering-climate feedback due to
seafloor weathering (Section \ref{sec:results}), it represents a
conservative assumption when considering the effect of changing the
land fraction on weathering. We discuss the effect of allowing a
seafloor weathering dependence on surface temperature in Section
\ref{sec:discussion}.

Combining Eqs.~(\ref{eq:weathering_nondim}), (\ref{eq:precip}), and
(\ref{eq:sf_weat2}), and assuming that continental weathering scales
linearly with land fraction ($\gamma$) and seafloor weathering scales
linearly with ocean fraction (1-$\gamma$), the total weathering rate
can be written
\begin{equation}
\label{eq:tot_weathering}
W=\beta_0 \frac{\gamma}{\gamma_0} (1+\alpha_pT)^a\phi^be^T + 
(1-\beta_0)\frac{1-\gamma}{1-\gamma_0} \left( \frac{\phi^n+\nu}{1+\nu} \right),
\end{equation}
where $\gamma_0$ is the modern Earth land fraction and
$\beta_0=\frac{W_{l0}}{W_{l0}+W_{o0}}$ is the fraction of total
weathering accomplished on continents for modern Earth conditions
($\gamma=\gamma_0$, $T=0$, and $\phi=1$).  $\beta_0$ will not
generally be the actual fraction of weathering accomplished on
continents for other climatic conditions. The magnitude of the
seafloor weathering rate on modern Earth is constrained by drilling
cores in the ocean floor, measuring the amount of basaltic rock that
has been weathered, and scaling to the entire ocean. Because
relatively few core holes have been drilled and analyzed, and because
weathering rates differ significantly among them, the seafloor
weathering rate on Earth is only roughly known
\citep{CALDEIRA:1995p2285}. \citet{CALDEIRA:1995p2285} assumes
$\beta_0=0.67$, \citet{Lehir08} assume $\beta_0=0.78$, and we will
take $\beta_0=0.75$ as our standard value, although we show that our
results are robust as long as $\beta_0 \approx 0.5-1$ in Section
\ref{sec:sensitivity}. Our assumption that continental and seafloor
weathering scale linearly with land and ocean fraction, respectively,
cannot be rigorously justified and should be viewed as a first
step. It is possible, for example, that most seafloor weathering
occurs near spreading centers so that total seafloor weathering is
nearly independent of ocean fraction. If true this would not affect
our main results significantly since they do not depend on the details
of the seafloor weathering parameterization (Section
\ref{sec:sensitivity}), but would affect an evaluation of the effect
of seafloor weathering on the outer edge of the habitable zone, where
strong seafloor weathering could keep CO$_2$ below the limit imposed
by CO$_2$ condensation or Rayleigh scattering (Section
\ref{sec:discussion}).

Global mean energy balance for a planet can be written
%
\begin{equation}
\label{eq:climate_model_dim}
\frac{\tilde{S}}{4}(1-\alpha)
= \tilde{\Omega}(\tilde{T},\tilde{\phi}),
\end{equation}
where $\tilde{S}$ is the dimensionful insolation ($S_0$=1365 W
m$^{-2}$ for modern Earth), $\alpha$ is the planetary Bond albedo, or
reflectivity ($\approx$$0.3$ for modern Earth), $\frac{1}{4}$ is a
geometric factor representing the ratio of the area of a planet
intercepting stellar radiation to the total planetary area, and
$\tilde{\Omega}(\tilde{T},\tilde{\phi})$ is the infrared radiation
emitted to space (outgoing longwave radiation). Below we will take
$\alpha=\alpha_0$ to be a constant for simplicity. We discuss this
assumption further in Section \ref{sec:discussion}.

After nondimensionalizing ($\Omega=\frac{\tilde{\Omega}}{S_0}$,
$S=\frac{S}{S_0}$), the emitted thermal radiation, $\Omega$, can be
Taylor expanded around current Earth conditions ($S=1$, $T=0$, and
$\phi=1$) to find
\begin{equation}
\label{eq:Omega_expansion}
\Omega(T,\phi)\approx\frac{1}{4}(1-\alpha)+\omega_1T-\omega_2\log \phi,
\end{equation}
where $\omega_1=\frac{\partial \Omega}{\partial T}|_{T=0,\phi=1,S=1}$
and $\omega_2=- \frac{\partial \Omega}{\partial \log
  (\phi)}|_{T=0,\phi=1,S=1}$. Note that for a planet with no
atmosphere, $\Omega\propto T^4$, as one would expect for a blackbody;
the current parameterization accounts for the effects of the
atmosphere, but is only a linearization of more complete radiative
transfer. Because the CO$_2$ absorption bands are saturated, the
outgoing longwave radiation is roughly linear in $\log \phi$
\citep{Pierrehumbert:2010-book}, so we expand in $\log \phi$. We
calculate $\omega_1$ and $\omega_2$ using published polynomial fits to
$\tilde{\Omega}$ calculated using a single column radiative-convective
model \citep{Pierrehumbert:2010-book}. We find $\omega_1$=0.015 using
$\omega_1=\tilde{\omega}_1 \frac{T_u}{S_0}$ and calculating
$\tilde{\omega}_1$ as the average slope of $\tilde{\Omega}$ between
$\tilde{T}=275$~C and $\tilde{T}=325$~K with CO$_2$=1000 ppm and a
relative humidity of 50\%.  We find $\omega_2$=0.007 using
$\omega_2=\frac{\tilde{\omega_2}}{S_0}$ and calculating
$\tilde{\omega_2}$ as the negative of the average slope of
$\tilde{\Omega}$ with respect to $\log \phi$ between CO$_2$=280 ppm
and CO$_2$=$2.8 \times 10^5$ ppm with a surface temperature of
$\tilde{T}$=280~K and a relative humidity of 50\%.  We show that our
main conclusions are unaltered if $\omega_1$ and $\omega_2$ are
changed by a factor or two in either direction in Section
\ref{sec:sensitivity}.  The linearization of $\Omega(T,\phi)$ leads to
the following climate model (nondimensional form of
Equation~(\ref{eq:climate_model_dim}))
\begin{equation}
\label{eq:climate_model}
\frac{1}{4}(S-1)(1-\alpha)=\omega_1T-\omega_2\log \phi.
\end{equation}

Taking $\Gamma$ to be the ratio of the CO$_2$ outgassing rate to that
on modern Earth, CO$_2$ equilibrium in the atmosphere requires
$\Gamma=W$, where $W$ is given by Equation~(\ref{eq:tot_weathering}). Such
an equilibrium is set up on a million year timescale, so that kinetic
(non-equilibrium) effects can be neglected when considering variations
in the weathering rate in response to changes in stellar luminosity.
We make order-of-magnitude estimates on a kinetic process in Section
\ref{sec:self-arrest} and we discuss the possibility that weathering
cannot increase enough to establish $\Gamma=W$ in Section
\ref{sec:discussion}. $\Gamma=W$, along with
Equation~(\ref{eq:climate_model}), represents a coupled climate-weathering
system that can be solved for temperature and CO$_2$ as a function of
insolation ($T(S)$ and $\phi(S)$). For simplicity we will take
$\Gamma=1$ in what follows, which is consistent with assuming a
constant seafloor spreading rate and plate tectonics.

\section{Effect of Land Fraction on Weathering}\label{sec:results}

\subsection{Simple Limits}
\label{sec:simple}

Throughout this paper we will be interested in the stability that the
carbon cycle can provide to the climate system. We will consider
specifically the change in surface temperature caused by a change in
insolation ($\frac{\partial T}{\partial S}$), which we will call the
climate sensitivity to external forcing.  The lower $\frac{\partial
  T}{\partial S}$ is, the more stable the climate is, the wider the
habitable zone will be, and the longer a planet will stay in the
habitable zone as its star evolves. We do not consider the many
effects that could become important near habitable zone limits, such
as water clouds on the inner edge
\citep{Selsis:2007p3441,Kaltenegger:2011p3443} and CO$_2$ clouds on
the outer edge
\citep{Forget:1997p3442,Selsis:2007p3441,Wordsworth:2011p3221,Forget2012,Wordsworth2012},
but $\frac{\partial T}{\partial S}$ will give us a sense of how wide
the habitable zone is in the broad range between where these effects
could be important.

We can calculate $\frac{\partial T}{\partial S}$ by differentiating
Equation~(\ref{eq:climate_model}) with respect to $S$
\begin{equation}
\label{eq:climate_model_diff}
\frac{\partial T}{\partial S} = \frac{1-\alpha}{4\omega_1}
+\frac{\omega_2}{\omega_1 \phi} \frac{\partial \phi}{\partial S}.
\end{equation}
The change in CO$_2$ as the insolation changes ($\frac{\partial
  \phi}{\partial S}$) defines the weathering feedback strength. In the
traditional picture of the silicate weathering feedback
\citep{Walker-Hays-Kasting-1981:negative}, the CO$_2$ decreases as the
insolation increases ($\frac{\partial \phi}{\partial S}<0$),
stabilizing the climate. If $\frac{\partial \phi}{\partial S}=0$ there
is no weathering feedback, and the temperature increases at a constant
rate, $\left(\frac{\partial T}{\partial S} \right)_0 =
\frac{1-\alpha}{4 \omega_1}$, as the insolation increases.

To calculate $\frac{\partial \phi}{\partial S}$, we need to consider
weathering. We can generalize the weathering relations we described in
Section \ref{sec:model} by rewriting Equation~(\ref{eq:tot_weathering}) as
\begin{equation}
\label{eq:tot_weathering_prime}
W=f(\gamma) W^\ast_l(\phi,T) + g(\gamma) W^\ast_o(\phi,T),
\end{equation}
where $f(\gamma)=\gamma$, $g(\gamma)=1-\gamma$,
$W^\ast_l(\phi,T)\equiv \frac{\beta_0}{\gamma_0}
(1+\alpha_pT)^a\phi^be^T$, and $W^\ast_o(\phi,T)\equiv
\frac{1-\beta_0}{1-\gamma_0} \left( \frac{\phi^n+\nu}{1+\nu} \right)$
in our standard model. The arguments in this section only require that
$f(\gamma=1)=g(\gamma=0)=1$ and $f(\gamma=0)=g(\gamma=1)=0$.

We will assume that the outgassing rate ($\Gamma$), and therefore the
total weathering rate ($\Gamma=W$), does not depend directly on
insolation ($S$). This is a reasonable assumption for surface
temperatures in the habitable zone, although the chemical composition
of outgassed material does vary at much higher temperatures
\citep{Schaefer:2010p3455}.  Using this assumption, we can
differentiate Equation~(\ref{eq:tot_weathering_prime}) with respect to $S$
to find
\begin{eqnarray}
 0=&&f(\gamma)
\left(\frac{\partial W^\ast_l}{\partial \phi} \frac{\partial \phi}{\partial S}+
\frac{\partial W^\ast_l}{\partial T} \frac{\partial T}{\partial S}\right)
\nonumber \\
&&+g(\gamma)
\left(\frac{\partial W^\ast_o}{\partial \phi} \frac{\partial \phi}{\partial S}+
\frac{\partial W^\ast_o}{\partial T} \frac{\partial T}{\partial S}\right).
\label{eq:weathering_diff}
\end{eqnarray}
Equation~(\ref{eq:weathering_diff}) can be solved for $\frac{\partial
  \phi}{\partial S}$, which can be substituted into
Equation~(\ref{eq:climate_model_diff}) to solve for $\frac{\partial
  T}{\partial S}$.

Before proceeding we consider two simple limits. First we consider a
waterworld ($\gamma=0 \to f=0, g=1$) in which seafloor weathering does
not depend on planetary surface temperature ($W^\ast_o(\phi,T) \equiv
W^\ast_o(\phi)$), as implied by most seafloor weathering
parameterizations (Section \ref{sec:model}). In this case all that is
left of Equation~(\ref{eq:weathering_diff}) is $\frac{\partial
  W^\ast_o}{\partial \phi} \frac{\partial \phi}{\partial S}=0$. If we
assume that $ W^\ast_o$ has a $\phi$ dependence so that it can adjust
to match outgassing, we are left with $\frac{\partial \phi}{\partial
  S}=0$. We therefore see that the assumption that seafloor weathering
does not depend on planetary surface temperature implies that there
can be no weathering feedback on a waterworld.

\begin{figure}[h!]
\begin{center}
  \includegraphics[width=20pc]{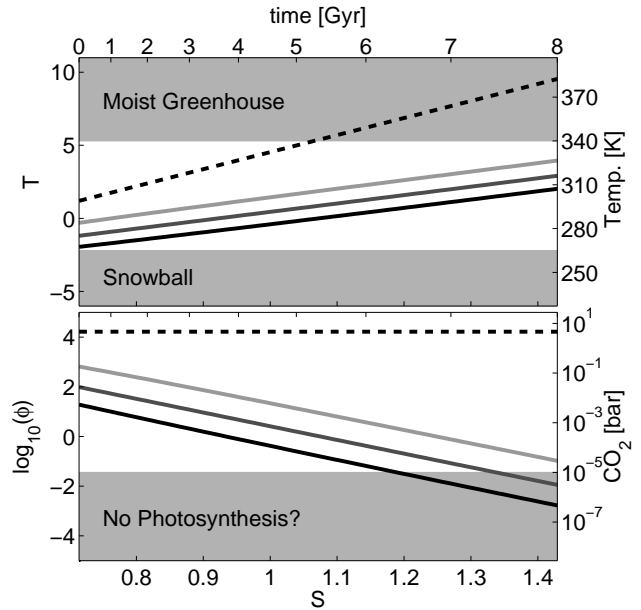}
\end{center}
\caption{Temperature and CO$_2$ concentration as a function of
  insolation with modern Earth volcanic outgassing ($\Gamma=1$) and
  various land fractions: $\gamma=0$ (dashed), $\gamma=0.01$ (light
  gray solid), $\gamma=0.1$ (dark gray solid), and $\gamma=0.99$
  (black solid). Both dimensionless (left) and dimensionful (right)
  values are given on axes. For reference, the time relative to zero
  age main sequence for a planet located 1~AU from a Sun-like star is
  plotted following \citet{GOUGH:1981p2371}. Gray shaded regions of
  the plot represent conditions that may be considered uninhabitable
  by some definition. The habitable zone upper temperature limit is
  the moist greenhouse, which we take to occur at 340~K ($T=5.3$)
  \citep{Kasting93}, and the lower temperature limit is global
  glaciation, which we take to occur at 265~K ($T=-2.2$)
  \citep{voigt10}, although it could occur 10-20~K colder than this
  \citep{Abbot-et-al-2011:Jormungand,yang2011a}. The lower limit
  placed on CO$_2$ for habitability in the figure, 10$^{-5}$ bar,
  \citep{CALDEIRA:1992p2366,VonBloh:2005p2325} should be understood
  only as a potential limit on complex photosynthetic life, and not on
  other types of life.}
\label{fig:evolve}
\end{figure}

As a second limit, we allow the land fraction ($\gamma$) to be
non-zero but make seafloor weathering ($W^\ast_o$) a constant
independent of temperature \emph{and} CO$_2$. While this limit does
not accurately describe our parameterizations in Section
\ref{sec:model}, we do expect seafloor weathering to be weaker than
continental weathering and to respond weakly to changes in CO$_2$
\citep{CALDEIRA:1995p2285}, which makes this limit useful for gaining
insight into the full model. Furthermore, seafloor weathering may be
limited by the availability of basaltic crust for reaction, which
would lead to constant seafloor weathering (Sections \ref{sec:model}
and \ref{sec:discussion}). If seafloor weathering is set constant,
then Equation~(\ref{eq:weathering_diff}) reduces to
\begin{equation}
\label{eq:weathering_diff2}
 0=f(\gamma)
\left(\frac{\partial W^\ast_l}{\partial \phi} \frac{\partial \phi}{\partial S}+
\frac{\partial W^\ast_l}{\partial T} \frac{\partial T}{\partial S}\right).
\end{equation}
Equation~(\ref{eq:weathering_diff2}) implies that the change in CO$_2$ as
insolation changes ($\frac{\partial \phi}{\partial S}$) is independent
of land fraction ($\gamma$) so that, by
Equation~(\ref{eq:climate_model_diff}), the climate sensitivity to external
forcing ($\frac{\partial T}{\partial S}$) must also be independent of
land fraction. This important result will lend insight to the full
model, in which we relax the assumption of constant seafloor
weathering, where we find that $\frac{\partial T}{\partial S}$ is
insensitive to land fraction ($\gamma$), even for very small land
fraction. Note that as long as $f(\gamma)$ is monotonically increasing
and all other parameters are held fixed, the continental weathering
rate decreases as the land fraction decreases
(Equation~(\ref{eq:tot_weathering_prime})). This would lead to increases in
temperature and CO$_2$ ($T$ and $\phi$) as land fraction ($\gamma$)
decreases, but they adjust precisely to keep $\frac{\partial
  T}{\partial S}$ constant in the limit of constant seafloor
weathering.

\subsection{Using the Full Model}

We now consider the full model described in Section \ref{sec:model}
rather than just the simple limits considered in Section
\ref{sec:simple}. This means, for example, that seafloor weathering is
given by Equation~(\ref{eq:sf_weat2}) rather than held constant. In the
full model there is a striking difference between the weathering
behavior of a waterworld and a planet with even a little land
(Figure~\ref{fig:evolve}). Varying the land fraction ($\gamma$) from
0.99 to 0.01 causes only a moderate warming and has a very small
effect on the climate sensitivity to external forcing ($\frac{\partial
  T}{\partial S}$). The functioning of the silicate weathering
feedback in the partially ocean-covered planets can be seen from the
decrease in CO$_2$ ($\phi$) with insolation ($S$;
Figure~\ref{fig:evolve}). Decreasing $\gamma$ from 0.99 to 0.01 would
therefore shift the habitable zone outward in insolation ($S$)
slightly, but would not significantly alter its width.

Setting the land fraction to zero ($\gamma=0$), however, causes a huge
shift in planetary weathering behavior. With $\gamma=0$, the CO$_2$
builds up to a high enough value that seafloor weathering can
accommodate outgassing, and stays at this high value regardless of
insolation ($S$), so there is no weathering-climate feedback. This
leads to a large increase in surface temperature, CO$_2$, and the
climate sensitivity to external forcing ($\frac{\partial T}{\partial
  S}$) for the waterworld relative to the partially ocean-covered
planets, so the waterworld crosses the moist greenhouse threshold much
sooner (Figure~\ref{fig:evolve}). Therefore a planet with even a small
amount of continent should enjoy a much wider habitable zone, and
longer tenure in it, than a waterworld.

None of the curves with $\gamma>0$ in Figure~\ref{fig:evolve} cross the
moist greenhouse threshold for $S<1.4$ due to the simplifications in
radiative transfer we have made, which is unrealistic when compared to
more detailed calculations \citep{Kasting93}; however, it is clear
that a planet with a lower land fraction will tend to be warmer and
experience a moist greenhouse at a lower insolation ($S$). On the
other hand, a lower land fraction ($\gamma$) leads to a higher CO$_2$
($\phi$), so a planet with a lower land fraction ($\gamma$) will cross
the CO$_2$ ($\phi$) threshold for C4 photosynthesis
\citep{VonBloh:2005p2325} at higher insolation ($S$).

\begin{figure}[h!]
\begin{center}
  \includegraphics[width=20pc]{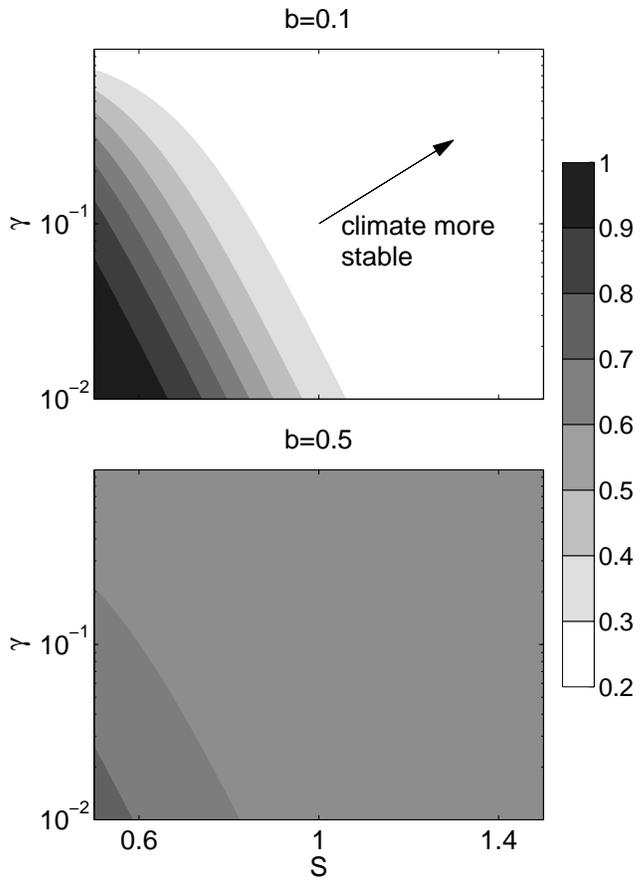}
\end{center}
\caption{Ratio of the climate sensitivity to external forcing
  ($\frac{\partial T}{\partial S}$) to the value it takes if there is
  no weathering feedback ($\left( \frac{\partial T}{\partial S}
  \right)_0=\frac{1-\alpha}{4 \omega_1}$), as a function of the
  insolation ($S$) and the land fraction ($\gamma$) with $b$=0.1
  (upper panel) and $b$=0.5 (lower panel). $b$ is the exponent of
  atmospheric CO$_2$ in the continental weathering rate
  parameterization (Equation~(\ref{eq:weathering_nondim})). Our
  standard value of $b$ is 0.5, and we consider a lower value because
  the effect of atmospheric CO$_2$ concentration on continental
  weathering may be reduced if land plants are present
  \citep{Pierrehumbert:2010-book}.}
\label{fig:dTdS_contour}
\end{figure}

We can solve for climate sensitivity to external forcing
($\frac{\partial T}{\partial S}$) explicitly as follows
\begin{equation}
\label{eq:dTdS_full}
\frac{\partial T}{\partial S} = \left( \frac{1-\alpha}{4 \omega_1} \right)
\left( \frac{b+\Theta(T,\phi,\gamma)}{\frac{\omega_2}{\omega_1}(1+\frac{a \alpha_p}{1+\alpha_p T})+
b+\Theta(T,\phi,\gamma)} \right),
\end{equation}
where
\begin{eqnarray}
\Theta(T,\phi,\gamma)=&&\left( \frac{(1-\beta_0)\gamma_0}{\beta_0 (1-\gamma_0)} \right)
\left( \frac{n(1-\gamma)}{\gamma} \right) \nonumber \\
&&\times \left(\frac{\phi^{n-b} e^{-T}}{
    (1+\nu) (1+\alpha_pT)^a} \right). \label{eq:Omega}
\end{eqnarray}
Note that the factor multiplying $\frac{1-\alpha}{4
  \omega_1}=\left(\frac{\partial T}{\partial S} \right)_0$ in
Equation~(\ref{eq:dTdS_full}) is always less than or equal to one, which
reduces the climate sensitivity to external forcing ($\frac{\partial
  T}{\partial S}$) below $\left(\frac{\partial T}{\partial S}
\right)_0$ and is a statement of the silicate weathering
feedback. Also, if $\beta_0=1$ (no seafloor weathering), $\Theta=0$
and $\frac{\partial T}{\partial S}$ has no dependence on the land
fraction ($\gamma$), as we expected from Section \ref{sec:simple}.
Additionally, $\lim_{\gamma, \beta_0 \to 0} \Theta = \infty$ and
$\lim_{\Theta \to \infty} \frac{\partial T}{\partial S} = \left(
  \frac{\partial T}{\partial S} \right)_0$, which recovers the
waterworld limit.

We plot the climate sensitivity to external forcing ($\frac{\partial
  T}{\partial S}$, Equation~(\ref{eq:dTdS_full})) as a function of
insolation ($S$) and land fraction ($\gamma$) for different values of
$b$ in Figure~\ref{fig:dTdS_contour}. It is useful to consider a lower
value of $b$ because the effect of CO$_2$ ($\phi$) on continental
weathering may be greatly reduced if land plants are present
\citep{Pierrehumbert:2010-book}, as we will discuss further in Section
\ref{sec:sensitivity}. The climate sensitivity to external forcing
($\frac{\partial T}{\partial S}$) increases for small land fraction
($\gamma$), but insolation ($S$) has a much stronger effect on
$\frac{\partial T}{\partial S}$ than land fraction ($\gamma$)
does. The climate sensitivity to external forcing ($\frac{\partial
  T}{\partial S}$) is highest at low insolation ($S$) and land
fraction ($\gamma$) and decreases as $S$ and $\gamma$
increase. $\lim_{\gamma \to 1} \Theta = 0$ and $\Theta \to 0$ as $S$
becomes large because $T$ becomes large (Equation~(\ref{eq:Omega})). This
allows the approximation that if either the planet is more than
$\approx$10\% land covered and/or the insolation is as large as or
greater than the modern Earth value, $\frac{\partial T}{\partial S}$
approaches the limit
\begin{eqnarray}
\frac{\partial T}{\partial S} &=& \left( \frac{\partial T}{\partial S} \right)_0
\left( \frac{b}{\frac{\omega_2}{\omega_1}(1+\frac{a
      \alpha_p}{1+\alpha_p T})+ b} \right) \nonumber \\
&\approx& \left(
  \frac{\partial T}{\partial S} \right)_0 \left(
  \frac{b}{\frac{\omega_2}{\omega_1}+ b}\right). \label{eq:dTdS_approx}
\end{eqnarray}
Therefore, despite the many parameters in our model (Section
\ref{sec:model}), the climate sensitivity to external forcing
($\frac{\partial T}{\partial S}$, Equation~(\ref{eq:dTdS_full})) can be
reduced to a function of only a few relevant variables over most of
the land fraction-insolation plane (Equation~(\ref{eq:dTdS_approx})). In
particular, if we neglect the (usually small) term $\frac{a
  \alpha_p}{1+\alpha_p T}$, the only weathering parameter present in
Equation~(\ref{eq:dTdS_approx}) is $b$, which controls the strength of the
scaling of land weathering with CO$_2$ (other remaining variables are
from the climate model).  The weaker land weathering scales with
CO$_2$ (lower $b$), the less $T$ can change while maintaining a
balance between outgassing and weathering, and the lower
$\frac{\partial T}{\partial S}$ is.  In the context of this paper,
what is important about Equation~(\ref{eq:dTdS_approx}) is that for a broad
range of insolation, the climate sensitivity to external forcing
($\frac{\partial T}{\partial S}$) is near the limit given by
Equation~(\ref{eq:dTdS_approx}) and is therefore nearly independent of land
fraction ($\gamma$). This explains the insensitivity of the climate
sensitivity to external forcing ($\frac{\partial T}{\partial S}$) to
land fraction in this model.

\section{Sensitivity to Model Parameters}
\label{sec:sensitivity}

\begin{figure*}[]
\begin{center}
  \includegraphics[width=40pc]{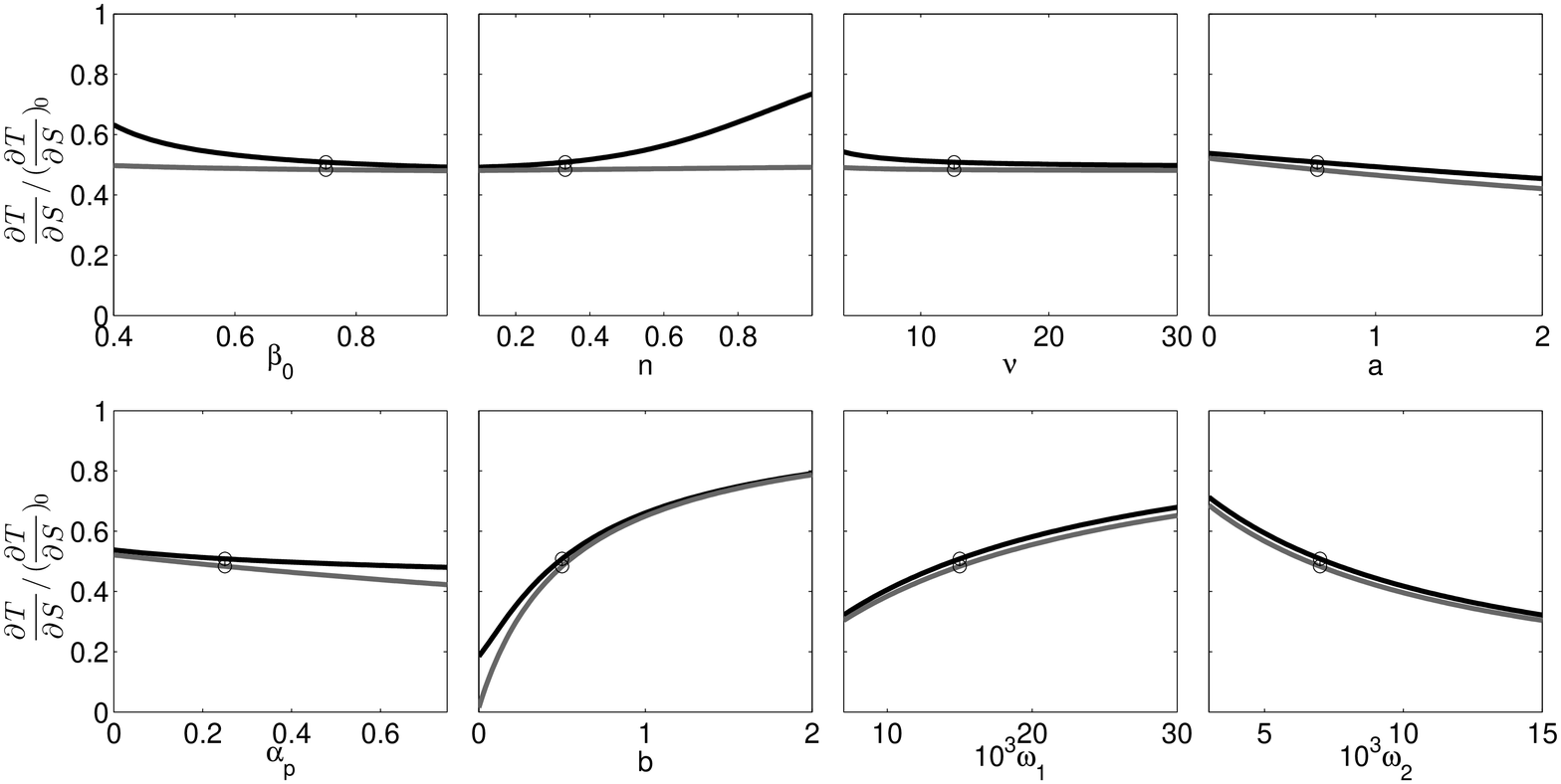}
\end{center}
\caption{Ratio of the climate sensitivity to external forcing
  ($\frac{\partial T}{\partial S}$) to the value it takes if there is
  no weathering feedback ($\left( \frac{\partial T}{\partial S}
  \right)_0=\frac{1-\alpha}{4 \omega_1}$) at both low land fraction
  ($\gamma=0.01$, thick black solid line) and moderate land fraction
  ($\gamma=0.3$, thick gray solid line). $\frac{\partial T}{\partial
    S}$ is displayed as a function of $\beta_0$, the fraction of total
  weathering accomplished on continents on Earth today; $n$, the
  exponent of atmospheric CO$_2$ ($\phi$) in the seafloor weathering
  parameterization (Equation~(\ref{eq:sf_weat2})); $\nu$, the constant
  in the seafloor weathering parameterization; $a$, the exponent of
  precipitation in the continental weathering parameterization
  (Equation~(\ref{eq:weathering_nondim})); $\alpha_p$, the constant
  defining the increase in precipitation with temperature
  (Equation~(\ref{eq:precip})); $b$, the exponent of atmospheric
  CO$_2$ ($\phi$) in the continental weathering parameterization
  (Equation~(\ref{eq:weathering_nondim})); $\omega_1$, the first term
  in the expansion of the outgoing longwave radiation ($\Omega$) with
  temperature ($\frac{\partial \Omega}{\partial T}$;
  Equation~(\ref{eq:Omega_expansion})); and $\omega_2$, the first term
  in the expansion of outgoing longwave radiation with the logarithm
  of CO$_2$ ($- \frac{\partial \Omega}{\partial \log (\phi)}$,
  Equation~(\ref{eq:Omega_expansion})). Circles and the thin black
  line signify standard values of parameters.}
\label{fig:vary_params}
\end{figure*}

In this Section we will evaluate the robustness of our result from
Section \ref{sec:results} that the weathering behavior, specifically
the climate sensitivity to external forcing ($\frac{\partial
  T}{\partial S}$), of a planet is relatively insensitive to surface
land fraction ($\gamma$). We will compare $\frac{\partial T}{\partial
  S}$ for a small land fraction ($\gamma=0.01$) with that for an
Earth-like land fraction ($\gamma=0.3$) while we vary model
parameters. For simplicity we consider reference insolation,
outgassing, and weathering values ($S=\Gamma=W=1$) and define
$\Delta=\left(\frac{\partial T}{\partial
    S}\right)\vert_{T=0,\phi=1,S=1,\gamma=0.01}-\left(\frac{\partial
    T}{\partial S}\right)\vert_{T=0,\phi=1,S=1,\gamma=0.3}$ as a
measure of the change in $\frac{\partial T}{\partial S}$ for a large
change in land fraction.

We first consider the effect on our results if current estimates of
the fraction of total weathering accomplished on continents on modern
Earth ($\beta_0$) were incorrect.  A lower $\beta_0$ would increase
$\frac{\partial T}{\partial S}$ (Equation~(\ref{eq:dTdS_full})), and this
effect is exaggerated when the land fraction ($\gamma$) is smaller,
increasing $\Delta$ (Figure~\ref{fig:vary_params}). For $\beta_0$ less
than approximately 0.5, $\Delta$ starts to become relatively
large. Given current estimates of 0.67 \citep{CALDEIRA:1995p2285} and
0.78 \citep{Lehir08} for $\beta_0$, seafloor weathering on modern
Earth would have to be 2-3 times larger than it is thought to be to
affect our main conclusions.

Next we consider changes in the seafloor weathering scheme. Ionic
concentration in seawater may either increase or decrease $n$
\citep{CALDEIRA:1995p2285}, the exponent of $\phi$ in the seafloor
weathering parameterization (Equation~(\ref{eq:sf_weat2})). The constant
$m$ in Equation~(\ref{eq:caldeira}), however, is typically taken to be less
than 1 \citep{CALDEIRA:1995p2285}, which would correspond to
$n=\frac{2}{3}$. Such a change would lead to a moderate increase in
$\Delta$ (Figure~\ref{fig:vary_params}), although a larger increase in
$n$ would significantly increase $\Delta$. It is less clear whether a
vastly different value of $\nu$ could be appropriate, but only an
extremely large decrease in $\nu$ would significantly increase
$\Delta$ (Figure~\ref{fig:vary_params}). The large effect that $\beta_0$
and $n$ would have on $\Delta$ if appropriate values were outside
current ranges highlights the importance of more experimental and
field work to better understand seafloor weathering.

The exponent of precipitation in Equation~(\ref{eq:weathering_nondim}),
$a$, and the constant defining the increase in precipitation with
temperature (Equation~(\ref{eq:precip})), $\alpha_p$, have a small effect
on $\frac{\partial T}{\partial S}$ (Figure~\ref{fig:vary_params}). This
is because the exponential temperature dependence of the Arrhenius
component of the continental weathering parameterization
(Equation~(\ref{eq:weathering_nondim})) dominates changes in continental
weathering due to temperature.  Moreover, changes in $a$ and
$\alpha_p$ affect $\frac{\partial T}{\partial S}$ similarly at low and
high land fraction ($\gamma$), leading to only very small changes in
$\Delta$ when these parameters are changed
(Figure~\ref{fig:vary_params}).

We next consider changes in $b$, the exponent of CO$_2$ in the
continental weathering parameterization. If seafloor weathering is
constant , setting $b=0$ yields a perfect thermostat ($\frac{\partial
  T}{\partial S}=0$), since weathering is a function of $T$ alone in
this case (Equation~(\ref{eq:tot_weathering})). As $b$ is decreased in the
full model, $\frac{\partial T}{\partial S}$ decreases, with larger
changes for larger land fraction ($\gamma$), since the model behaves
more like the constant seafloor weathering case in this limit
(Figure~\ref{fig:vary_params}). For very small $b$, $\Delta$ becomes
somewhat large, although this effect is greatly diminished for
insolation larger than the reference value $S>1$
(Figure~\ref{fig:dTdS_contour}), when the temperature becomes large
enough to make $\Theta$ small (Section \ref{sec:results}).

Finally, we consider the terms defining the expansion of infrared
emission to space ($\Omega$, outgoing longwave radiation) as a
function of temperature ($\omega_1=\frac{\partial \Omega}{\partial
  T}$) and the logarithm of CO$_2$ ($\omega_2=- \frac{\partial
  \Omega}{\partial \log (\phi)}$). We only kept the first order terms
in these expansions, and we made specific assumptions to calculate
them (Section~\ref{sec:model}), so it is worth considering whether
variations in $\omega_1$ and $\omega_2$ affect our results. Decreasing
$\omega_1$ has the simplistic effect of increasing the climate
sensitivity to external forcing with no weathering feedback $\left(
  \frac{\partial T}{\partial S} \right)_0=\frac{1-\alpha}{4
  \omega_1}$. We focus on weathering feedbacks in
Figure~\ref{fig:vary_params} by normalizing the climate sensitivity to
external forcing by $\left( \frac{\partial T}{\partial S} \right)_0$.
$\omega_1$ and $\omega_2$ enter the relation for the weathering
feedback in Equation~(\ref{eq:dTdS_full}) (the part multiplying $\left(
  \frac{\partial T}{\partial S} \right)_0$ on the right hand side)
through the ratio $\frac{\omega_2}{\omega_1}$, which is a measure of
the climate sensitivity to changes in CO$_2$. Increasing the climate
sensitivity to changes in CO$_2$ ($\frac{\omega_2}{\omega_1}$)
increases the warming effect of a given increase in CO$_2$, increasing
the weathering feedback strength, and decreasing the climate
sensitivity to external forcing (Figure~\ref{fig:vary_params}). This
effect shows little dependence on land fraction, so large changes in
$\omega_1$ and $\omega_2$ do not alter our main results
(Figure~\ref{fig:vary_params}).

We note that $T_u$, the exponential temperature scale for the direct
temperature dependence in Equation~(\ref{eq:weathering}), comes into the
nondimensional relations through $\alpha_p=\tilde{\alpha}_pT_u$ and
$\omega_1=\tilde{\omega}_1 \frac{T_u}{S_0}$. Since reasonable
variations in $\alpha_p$ and $\omega_1$ do not alter our main
conclusions, we can conclude that reasonable changes in $T_u$ would
not. Indeed, we find that the climate sensitivity to external forcing
depends minimally on land fraction ($\gamma$) when we vary $T_u$
between 5 and 25~K (not shown).

\section{Waterworld Self-Arrest}\label{sec:self-arrest}

\begin{figure}[h!]
\begin{center}
  \includegraphics[width=20pc]{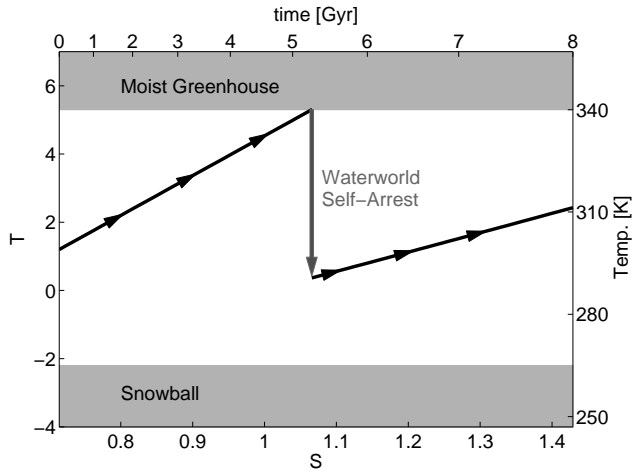}
\end{center}
\caption{Schematic diagram showing how the waterworld from
  Figure~\ref{fig:evolve} could experience a waterworld self-arrest as
  the insolation ($S$) increases and end up as an Earth-like planet
  with partial ocean coverage ($\gamma=0.3$ in this case) in the
  habitable zone. Axes are as in Figure~\ref{fig:evolve}.}
\label{fig:self_arrest}
\end{figure}

Due to the lack of a climate-weathering feedback in a waterworld, we
have argued that a waterworld should experience a moist greenhouse at
much lower insolation ($S$) and have a smaller habitable zone than a
planet with even a very small land fraction. We will now discuss the
possibility that a waterworld could stop a moist greenhouse by
engaging silicate weathering when enough ocean is lost that continent
is exposed, engaging weathering and reducing temperature ($T$), and
thereby preventing complete water loss. We will refer to this
possibility as a ``waterworld self-arrest'' (see
Figure~\ref{fig:self_arrest} for a schematic diagram of the idea). The
waterworld self-arrest is similar to the idea, discussed by
\citet{abe2011}, that a planet with partial ocean coverage could
experience a moist greenhouse and become a habitable land planet.

\begin{figure}[h!]
\begin{center}
  \includegraphics[width=20pc]{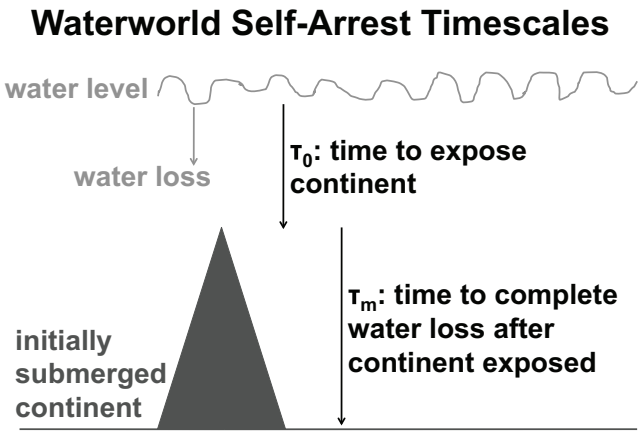}
\end{center}
\caption{Schematic diagram of the timescales involved in a waterworld
  self-arrest. $\tau_0$ is the time the planet exists in a moist
  greenhouse state before submerged continent is exposed. $\tau_m$ is
  the time to complete planetary water loss after the first exposure
  of continent. In order to determine whether a waterworld self-arrest
  is possible, $\tau_m$ must be compared with $\tau_w$, the timescale
  for CO$_2$ to be sufficiently reduced by silicate weathering that
  moist greenhouse conditions cease to exist.}
\label{fig:self_arrest_diagram}
\end{figure}

If a waterworld experiences a moist greenhouse, there will be some
amount of time before any continent is exposed and continental
weathering is possible. We call this timescale $\tau_0$
(Figure~\ref{fig:self_arrest_diagram}), and we will not use it in the
present argument. We call the time after the first exposure of
continent until all remaining water is lost through hydrogen escape to
space $\tau_m$. To determine whether a waterworld self-arrest is
possible, we must compare $\tau_m$ to the timescale for CO$_2$ to be
reduced by continental silicate weathering to a level such that moist
greenhouse conditions cease to exist, which we call $\tau_w$. If
$\tau_w<\tau_m$, then a waterworld self-arrest is possible.

If we take the hydrogen atom mixing ratio in the stratosphere to be
unity as a limiting case, the diffusion-limited rate of loss of
hydrogen atoms to space is $2\times 10^{13}$ H atoms cm$^{-2}$
s$^{-1}$ \citep{Kasting93}. Earth would be essentially a waterworld if
the ocean were a few times more deep, which would correspond to
$\approx 10^{29}$ H atoms cm$^{-2}$, so that $\tau_m$ is of order 100
Myr. Since we seek only an order of magnitude here, we do not consider
the effect of different topography on $\tau_m$, but note that larger
planets will tend to have less topography, which would reduce $\tau_m$
somewhat. We note also that energetic limitation on hydrogen escape
could increase $\tau_m$ \citep{Watson81,abe2011}.

$\tau_w$ is much more difficult to estimate than $\tau_m$. The closest
known analog to a moist greenhouse in post-Hadean Earth is likely the
aftermath of the Snowball Earth events, which occurred 600--700
million years ago. The dominant explanation for Snowball Earth events
\citep{Kirschvink92,Hoffman98} is that some cooling factor forced
Earth into a global glaciation, which is an alternative stable state
of climate because of the high reflectivity of snow and ice. Because
it would be very cold during a Snowball and because any precipitation
would fall as snow, rather than rain, the continental weathering rate
would be near zero. As we expect volcanic CO$_2$ outgassing to
continue unabated, the atmospheric CO$_2$ would increase until it
reached up to a thousand times its present value
($\phi=\cal{O}$(10$^3$)), at which point it would be warm enough to
melt the ice and end the Snowball.  In the immediate aftermath of a
Snowball Earth event the planetary albedo would be very low and the
CO$_2$ concentration would be very high, which would lead to surface
temperatures in the range of a moist greenhouse. Extremely high
weathering would then reduce the atmospheric CO$_2$ until Earth's
surface temperatures were comparable to modern values
\citep{Higgins03}. If post-Snowball Earth was truly in a moist
greenhouse state, complete water loss must not have occurred, which
would imply that $\tau_w<\tau_m$ and a waterworld self-arrest is
possible. Even if post-Snowball Earth was not a true moist greenhouse,
the drawdown of CO$_2$ must have been immense, and geochronological
evidence indicates that this took less than 4 Myr
\citep{Condon:2005p3059}. This further motivates the idea that
$\tau_w$ could be small compared to $\tau_m$.

The CO$_2$ required to end a Snowball Earth, however, may have been
1-2 orders of magnitude lower than that quoted above
\citep{Abbot-Pierrehumbert-2009:mudball}, in which case the Snowball
aftermath would be a weaker analog for a moist greenhouse and further
consideration is required.  Geochemical modeling under conditions
similar to a moist greenhouse yield weathering rates of
$\cal{O}$($10^{13}$) kg C yr$^{-1}$ \citep{Higgins03} and
$\cal{O}$($10^{12}$) kg C yr$^{-1}$ \citep{LeHir:2009p898}, although
\citet{mills11} suggest that ion transport constraints could limit the
weathering rate to $\approx 5 \times 10^{11}$ kg C yr$^{-1}$
\citep{mills11}.  The waterworld CO$_2$ is $1.65 \times 10^4$ times
the present value (Figure~\ref{fig:evolve}) in our model with standard
parameters, which corresponds to $\cal{O}$($10^{19}$ kg C).  This
yields an estimate of $\tau_w=\cal{O}$(1-10 Myr), which further
suggests that $\tau_w<\tau_m$.

Although we have not performed a full analysis of the kinetic
(non-equilibrium) effects, the order-of-magnitude analysis we have
done indicates that a habitable zone waterworld could stop a moist
greenhouse through weathering and become a habitable partially
ocean-covered planet. We note that this process would not occur if the
initial water complement of the planet is so large that continent is
not exposed even after billions of years in the moist greenhouse state
($\tau_0>\cal{O}$(10 Gyr)). Additionally, we have neglected the
possibility of water exchange between the mantle and the surface,
which is poorly constrained on Earth. If, for some reason, water were
continuously released from the planetary interior to replace water
lost through hydrogen escape, then the conclusions reached here would
be invalid. This would require significantly higher mantle-surface
fluxes than on present Earth, which are constrained by cerium
measurements at mid-ocean ridges to be small enough to require
$\approx$10~Gyr to replace Earth's surface water reservoir
\citep{Hirschmann:2012p3481}.  Finally, we note that since we have
considered a moist greenhouse caused by high atmospheric optical
thickness rather than high insolation, the waterworld would not be
likely to suffer a full runaway greenhouse during the process
described here \citep{Nakajima92}.

\section{Discussion of Observability}\label{sec:observables}

The theory developed here suggests that if an Earth-size planet is
detected near the habitable zone it should be considered a good
candidate to have a wide habitable zone and a long-term stable climate
as long as it has some exposed land. The first observational point to
consider, therefore, is whether the surface land fraction could be
measured on an Earth-like planet.

Using measurements of the rotational variations of disk-integrated
reflected visible light from a future mission such as TPF-C it will be
possible to determine the surface land fraction of an Earth-like
planet orbiting a Sun-like star. \citet{Cowan:2009p3255} used
observations of Earth from the Deep Impact spacecraft to demonstrate
that this is possible with multi-band high-cadence photometry.  The
multi-band observations are critical because the albedo of land and
ocean have very different wavelength dependencies, while the
integration times must be sufficiently short to resolve the planet's
rotational variability. Clouds tend to obscure surface features and
make such a determination more difficult
\citep{Ford:2001p2896,Kaltenegger:2007p3446,Palle:2008p2895}, and
regions of a planet that are consistently cloudy are impervious to
this form of remote sensing
\citep{Cowan:2011p3304}. \citet{Kawahara:2010p3262,Kawahara:2011p3261}
have extended this method to include seasonal changes in illumination
to generate rough two-dimensional maps of otherwise unresolved
planets. Provided that a habitable zone planet is not much cloudier
than Earth, its land fraction should therefore be observationally
accessible with sufficient signal-to-noise ratio, and the present
study can be used to help assess the probability of its habitable zone
width and long-term habitability.

The presence of exposed land could be detected on an exoplanet via an
infrared emission spectrum, provided the overlying atmosphere is not
too much of a hindrance \citep{Hu_2012}. At thermal wavelengths, the
land fraction of a planet could in principle be estimated based on its
thermal inertia, but in idealized tests \citet{Gaidos:2004p2921} found
that the obliquity-inertia degeneracy impedes quantitative estimates
of a planet's heat capacity. It is therefore unlikely that a thermal
mission, on its own, could constrain the land fraction of an
exoplanet.

Direct confirmation or falsification of the theory developed here will
be more difficult, requiring not only an estimate of land fraction,
but also atmospheric CO$_2$ abundance and/or surface temperature. The
difference between the temperature and CO$_2$ of a waterworld and a
partially ocean-covered planet could be quite large, particularly near
the inner (hotter) edge of the habitable zone
(Figure~\ref{fig:evolve}). This entails measuring, in exoplanets,
discrepancies in carbon dioxide abundance of roughly four orders of
magnitude, and surface temperatures differences of $\sim 70$~K.

Detecting carbon dioxide in an exoplanet atmosphere is relatively
easy: the \textit{James Webb Space Telescope} will determine whether
CO$_2$ is present in the atmospheres of transiting super-Earths
orbiting M-stars
\citep{Deming:2009p3264,Kaltenegger:2009p3447,Belu:2011p3449,Rauer:2011p3448}. The
challenge is in estimating the \emph{amount} of CO$_2$ in an
atmosphere. In order to measure the carbon dioxide \emph{abundance} of
an atmosphere, the atmospheric mass must be estimated.  This could be
achieved for M-Dwarf habitable zone planets with transit spectroscopy
covering the Rayleigh scattering slope \citep{Benneke_2012}. The
unknown radius of directly-imaged (longer-period) exoplanets may
hinder such estimates with a TPF-C type mission. At thermal
wavelengths, the planetary radius will be measurable and the pressure
broadening of absorption features may offer an estimate of surface
pressure and hence atmospheric mass
\citep[e.g.,][]{DesMarais:2002p3451,Meadows-Seager:2010}.

Using a thermal mission like TPF-I, it may be possible to determine
the surface temperature of an Earth analog via its brightness
temperature in molecular opacity windows, at least for a cloud-free
atmosphere with few IR absorbers
\citep{DesMarais:2002p3451,Kaltenegger:2010p3454,Selsis:2011p3305}. While
low-lying clouds do not significantly degrade such estimates, high
cloud coverage can lead to surface temperature estimates at least 35~K
too low \citep{Kaltenegger:2007p3446,Kitzmann:2011p3306}.  Continuum
opacity from water vapor will also lead to lower brightness
temperatures, even in opacity windows.  This could lead to
under-estimating the surface temperature of planets with high vapor
content, precisely the moist greenhouse planets expected to have high
surface temperatures. Moreover, Venus dramatically demonstrates that
brightness temperatures, even in supposed windows, are not necessarily
a good probe of the surface temperature.

Given the similar temperature and CO$_2$ among partially ocean-covered
planets of varying land fraction (Figure~\ref{fig:evolve}), it is
unlikely that any mission in the foreseeable future would be able to
distinguish between them. Additionally, factors that we have neglected
here introduce uncertainties that make the temperature and CO$_2$
abundance among partially ocean-covered planets effectively
indistinguishable.

To summarize, measurements from a mission like TPF-C would provide a
surface land fraction estimate for planets not much cloudier than
Earth. TPF-C could provide an order-of-magnitude atmospheric CO$_2$
estimate if the radius were known, which would be possible if (1) the
planet were transiting, (2) its radius were estimated from TPF-I
observations, or (3) its radius were estimated based on a mass
measurement. TPF-I could provide a surface temperature estimate that
would likely be good to $\cal{O}$(10~K) for an Earth analog, but
possibly much worse for a cloudier and/or more humid atmosphere. It
would therefore be possible to directly test our predictions for
differences in weathering behavior of waterworlds and partially
ocean-covered planets if data from missions like both TPF-C and TPF-I
were available, and sufficient numbers of nearby habitable zone
planets exist. Prospects are limited, but testability is still
possible in special cases (such as a transiting planet), if only one
type of data is available.

A final observational point is that in Section \ref{sec:self-arrest}
we calculated that a waterworld self-arrest, through which a fully
ocean planet would be transformed into a partial ocean planet, is
possible. Waterworlds, however, would still exist and their detection
would not falsify the waterworld self-arrest idea. Waterworlds could
have orders of magnitude more water than Earth
\citep[e.g.,][]{Raymond:2007,Fu2010}, and could therefore exist in a
moist greenhouse state for many billions of years. Additionally, a
waterworld in the outer regions of the habitable zone would not be in
a moist greenhouse state (Figure~\ref{fig:self_arrest}).

\section{General Discussion}\label{sec:discussion}

We note again that we make numerous approximations in our model, so
that it is useful for qualitative understanding rather than
quantitative predictions. Our linearized climate model, for example,
is a gross simplification of the real climate system, which is
probably responsible for the delayed onset of the moist greenhouse in
our model as the insolation is increased (Figure~\ref{fig:evolve})
relative to more detailed calculations \citep{Kasting93}. Our
approach, however, is appropriate for a general study of potential
terrestrial exoplanets, whose detailed characteristics are not known,
and our main conclusions should be robust to more detailed
modeling. For example, we point out that no weathering feedback is
possible ($\frac{\partial T}{\partial S}=\left( \frac{\partial
    T}{\partial S} \right)_0$) for a waterworld as long as seafloor
weathering is independent of surface temperature ($T$), regardless of
the specific form of the seafloor weathering parameterization.

Nonetheless, our parameterization of seafloor weathering is probably
the most uncertain aspect of this work. For example, we have neglected
the potential effect of a different ocean chemical composition on an
exoplanet, which could lead to different base weathering rates or a
different scaling of seafloor weathering with CO$_2$
($\phi$). Furthermore, in order to derive a scaling relationship
between [H$^+$] and $\phi$, we have approximated ocean chemistry as a
carbonate ion system with changes in alkalinity determined by changes
in Ca$^{2+}$.  Real oceans have other ions present, which can alter
the charge balance. These approximations, however, have been useful
for modeling carbon cycling in deep time Earth problems
\citep[e.g.,][]{Higgins03} and our knowledge of typical ion
concentrations in terrestrial exoplanets is unconstrained. Finally, we
have assumed that surface temperature has a negligible effect on
seafloor weathering rate, which we will now consider in more detail.

As there is a temperature dependence of the reactions relevant for
seafloor weathering \citep{Brady:1997p3530}, we have effectively
assumed that surface temperature does not strongly affect the
temperature at which seafloor weathering occurs. If a dependence of
seafloor weathering on surface temperature is allowed, a
climate-seafloor weathering feedback is possible, which should tend to
reduce the effect of changing land fraction. In terms of the effect of
land fraction on weathering behavior, we have therefore considered a
conservative case in assuming no surface temperature dependence of
seafloor weathering, yet we still found that land fraction has only a
small effect on weathering behavior for partially ocean-covered
planets.

A strong temperature dependence of seafloor weathering would affect
waterworld behavior at the inner edge of the habitable zone. We can
see this by multiplying the seafloor weathering parameterization
(Equation~(\ref{eq:sf_weat2})) by an exponential temperature dependence of
the form $e^{\eta T}$, where $\eta=\frac{d T_u}{T_w}$, where $d$ is
the change in seafloor weathering reaction temperature per unit change
in surface temperature, and $T_w \approx 15$K is the exponential scale
of increase in seafloor weathering reaction rate with temperature
\citep{Brady:1997p3530}.  The largest value $\eta$ could take would be
if current reactions occur near the bottom water temperature
($\approx$0$^\circ$C) and each unit increase in surface temperature
creates a unit increase in bottom water temperature ($d$=1,
$\eta=\frac{2}{3}$).  Our standard assumption is that seafloor
weathering reactions occur at a temperature of $\approx$20-40$^\circ$C
regardless of surface temperature, which is equivalent to
$d$=$\eta$=0. Using $\eta=\frac{2}{3}$ increases the insolation
required to cause a waterworld moist greenhouse from $S$$\approx$1.2 to
$S$$\approx$2.1 if $\nu$=0 and prevents a waterworld moist greenhouse
from ever happening if $\nu$=12.6. A strong temperature dependence of
seafloor weathering would therefore make it much harder for a
waterworld to experience a moist greenhouse, although we do not
consider the limiting case considered in this paragraph likely.

As noted in Section \ref{sec:model}, we assume that seafloor spreading
produces a sufficient quantity of weatherable material and a
sufficient amount of water can circulate through it that the seafloor
weathering rate can increase arbitrarily as the ocean pH
decreases. Since we assume seafloor weathering accounts for 25\% of
weathering on modern Earth ($\beta_0=0.75$), if the CO$_2$ outgassing
rate takes its modern value ($\Gamma=1$), then the seafloor weathering
rate must be able to increase by a factor of four if it alone is to
balance outgassing, as would be required to establish a CO$_2$ steady
state on a waterworld. If there is not enough weatherable material for
this to be possible, for example due to weak circulation of seawater
through weatherable material, then the waterworld CO$_2$ will not be
limited to the value calculated in Section \ref{sec:results}
(Figure~\ref{fig:evolve}), but will instead steadily increase due to an
unbalanced source and sink. This would push a waterworld more quickly
to a moist greenhouse, and potentially a waterworld self-arrest, than
one would otherwise expect. A related issue is that if the land
fraction is very small, there may be physical limits on continental
weathering not included in Equation~(\ref{eq:weathering_nondim}). This
would lead to higher CO$_2$ concentrations and temperatures for small
land fractions than those found in Section \ref{sec:results}. The
smallest land fraction we explicitly considered, however, was
$\gamma=0.01$, which corresponds to the combined area of Greenland and
Mexico on Earth. Given continuous resurfacing it is not clear that
physical limitations on weathering would be in play on such a large
land area.

\begin{figure}[h!]
\begin{center}
  \includegraphics[width=20pc]{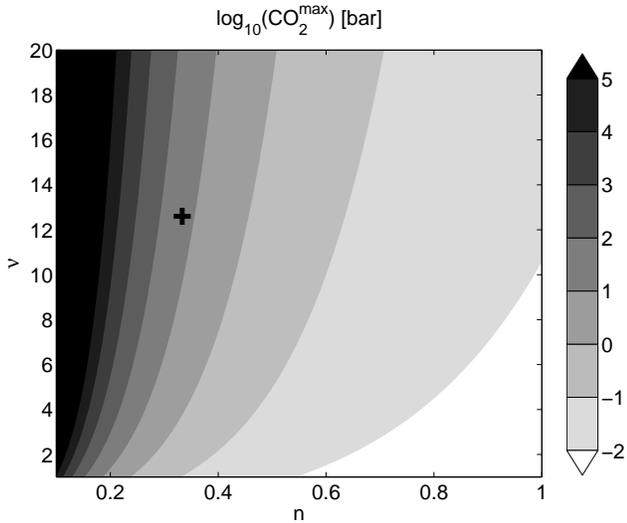}
\end{center}
\caption{Maximum atmospheric partial pressure of CO$_2$ (dimensionful
  version of $\bar{\phi}$) allowed by seafloor weathering, as a
  function of $n$ and $\nu$, parameters in the seafloor weathering
  scheme (Equation~(\ref{eq:sf_weat2})). The ``+'' represents standard
  parameter values. }
\label{fig:max_CO2}
\end{figure}

We have focused on the effect on land fraction on weathering behavior
relevant for the broad interior of the habitable zone and the
evolution of a planet through the habitable zone over time.  The
effect of land fraction on the outer (cold) edge of the habitable zone
is of interest as well. Based on increased Rayleigh scattering or
CO$_2$ condensation at the surface, the outer edge of the habitable
zone occurs when CO$_2$$\approx$5--10~bar
\citep{Kasting93}. \citet{Lehir08}, however, found that seafloor
weathering should limit the maximum CO$_2$ concentration on Earth to a
few tenths of a bar. As pointed out by \citet{Pierrehumbert:2011p3287},
if this is the case, then the CO$_2$ condensation limit could not be
reached and seafloor weathering would shrink the habitable zone. Using
our seafloor weathering parameterization, we can derive an upper limit
on the atmospheric CO$_2$ concentration with Earth's land fraction
($\gamma=\gamma_0$): $\bar{\phi}=\left(\frac{\Gamma (1+\nu)
  }{(1-\beta_0)} - \nu\right)^\frac{1}{n}$. Using our standard
parameters, we find $\bar{\phi}=7.3\times10^4$, or $\approx$20 bar.
$\bar{\phi}$ depends strongly, however, on model parameters
(Figure~\ref{fig:max_CO2}), particularly $n$. $\bar{\phi}=10^3$, roughly
the value \citet{Lehir08} found, would be produced by our model by
either a change in $n$ from 0.33 to 0.54 or a change in $\nu$ from
12.6 to 2.0. Therefore, in contrast to our main results (Section
\ref{sec:sensitivity}), the weathering behavior relevant for the outer
edge of the habitable zone is highly sensitive to the seafloor
weathering parameterization. A convincing study of the effect of
seafloor weathering on the outer edge of the habitable zone will have
to wait until seafloor weathering is better understood.

We have found that relatively ineffective seafloor weathering will
tend to lead to much warmer temperatures in a waterworld than a
partial ocean planet. This might appear to contrast with
\citet{Sleep:2001p3368}, who calculated very cold temperatures for
Earth during the Hadean, when Earth had a much lower land fraction and
was further out in the habitable zone. The primary reason for the
difference is that \citet{Sleep:2001p3368} include a significant sink
of carbon due to weathering of impact ejecta during the Hadean, which
significantly reduces the equilibrated CO$_2$ and temperature. They
also assume that seafloor weathering is much more active due to
increased internal heat flux early in Earth's history. We do not
include either of these effects because our effort is aimed at general
consideration of the problem in the context of exoplanets, but we
acknowledge that either could limit the CO$_2$, and therefore the
probability of a moist greenhouse, on a waterworld. Finally,
\citet{Sleep:2001p3368} also find a strong dependence of equilibrated
CO$_2$ on the strength of the CO$_2$ dependence of seafloor
weathering, as we do (Figure~\ref{fig:max_CO2}), early in Earth's
history when the CO$_2$ may have been very high. This again emphasizes
that the details of the seafloor weathering scheme are critical for
its effect on the outer edge of the habitable zone, or equivalently
the climate of early Earth.

In our analysis we set the planetary albedo to a constant for
simplicity. Since the albedo of land
\citep[0.05-0.3,][]{Peixoto-Oort-1992:physics} tends to be higher than
the albedo of ocean \citep[0.05-0.1,][]{Peixoto-Oort-1992:physics},
one would expect the planetary albedo to increase as the land fraction
increases. This would decrease the temperature (T,
Equation~(\ref{eq:climate_model})) and the climate sensitivity to external
forcing ($\frac{\partial T}{\partial S}$, Equation~(\ref{eq:dTdS_full})),
both of which we found would already be lower due to more continental
weathering (Section \ref{sec:results}). This effect would likely not
be very large, however, as detailed analysis of satellite measurements
indicates that because of extensive absorption and reflection by
clouds and gas in the atmosphere the surface contributes only
$\cal{O}$(10\%) to the global mean albedo on Earth
\citep{Donohoe:2011p3253}. The influence of surface albedo on
planetary albedo would be further reduced around M stars, which emit
at longer wavelengths, due to increased atmospheric scattering and
absorption \citep{Kasting93}. Another surface albedo effect we have
neglected is the change in albedo in colder climates due to the
formation of ice and snow at the surface. This would lead to a
reduction in the range of $S$ for which a planet is habitable in
Figure~\ref{fig:evolve}. Again, however, this effect would be reduced
around M stars due to greatly reduced ice and snow albedos
\citep{Joshi11}.

Our model neglects spatial variation in climate and weathering, and
therefore cannot address spatially heterogeneous effects such as the
feedbacks described by \citet{Kite:2011}. An assumption of spatial
homogeneity, however, is a reasonable first approximation even on
planets with highly asymmetric insolation distributions, such as
tidally locked planets in the habitable zone of M stars, if there is
an ocean or atmosphere similar in size to those on Earth to
redistribute heat
\citep[e.g.,][]{Joshi:1997,Merlis:2010,Edson:2011p2367}.

Current understanding is that continents were not present at Earth's
formation, but started to form soon afterward. For example, zircon
evidence indicates that there was continental crust 4.4 Gyr ago, or
within $\approx$100~Myr of Earth's formation
\citep{Wilde:2001p3529}. Available evidence cannot constrain whether
the continents reached their present size quickly (within
$\approx$500~Myr) or have grown slowly over Earth's history and continue
to grow today \citep{Eriksson:2006p3498}. We have neglected such
processes, but this is unlikely to modify our main conclusions because
we found that the climate sensitivity to external forcing
($\frac{\partial T}{\partial S}$) does not depend strongly on land
fraction. Our work suggests that the order of magnitude of water
delivery, which determines whether a planet is dry, partially
ocean-covered, or a waterworld, is much more important for determining
weathering behavior than evolution in the size of continents through a
planet's lifetime. Finally, it is worth speculating that Earth could
have experienced a waterworld self-arrest early in its history,
although much more thought is needed to determine what the geochemical
signals left behind by such an event would be.

If a waterworld has a thick enough ocean, the overburden pressure
could significantly suppress volcanic outgassing. For example, CO$_2$
degassing could be completely suppressed on an Earth-like planet with
100~km of ocean \citep{Kite:2009p2923}. This uncertainty is
commensurate with other volcanic outgassing uncertainties, such as
CO$_2$ content in the magma, for partial ocean planets and waterworlds
that barely cover their continents with ocean. Overburden outgassing
suppression could lead to lower CO$_2$ levels and higher chance of a
Snowball state than predicted here for waterworlds with a thick ocean
layer.

In our discussion of a waterworld self-arrest (Section
\ref{sec:self-arrest}) we neglected effects of changing planetary
size. Super-Earths would have higher gravity and therefore reduced
orography.  This will reduce the amount of time it takes from first
exposure of land to complete water loss ($\tau_m$), which will make a
waterworld self-arrest less likely. On the other hand, the higher
gravity of a super-Earth will also make gas escape from the atmosphere
to space more difficult, which would reduce $\tau_m$. Full
consideration of a waterworld self-arrest on a super-Earth is beyond
the scope of this paper.

\section{Conclusions}\label{sec:conclusions}

We have shown that using standard weathering parameterizations, the
weathering behavior of a partially ocean-covered Earth-like planet
does not depend strongly on land fraction, as long as the land
fraction is greater than $\approx$0.01. Consequently, planets with
some continent and some ocean should have a habitable zone of similar
width. This is a powerful result because it indicates that previous
habitable zone theory developed assuming Earth-like land fraction and
weathering behavior should be broadly applicable.

We have also pointed out that, as long as seafloor weathering does not
depend directly on surface temperature, a climate-weathering feedback
cannot operate on a waterworld. This is a significant result because
it would imply that waterworlds have a much narrower habitable zone
than a planet with even a few small continents. We find, however, that
weathering could operate quickly enough that a waterworld could
``self-arrest'' while undergoing a moist greenhouse and the planet
would be left with partial ocean coverage and a clement climate. If
this result holds up to more detailed kinetic weathering modeling, it
would be profound, because it implies that waterworlds that form in
the habitable zone have a pathway to evolve into a planet with partial
ocean coverage that is more resistant to changes in stellar
luminosity.

\section{Acknowledgements}
We thank Jens Teiser for suggesting that we consider whether
weathering could draw down CO$_2$ fast enough to stop a moist
greenhouse and David Archer and Ray Pierrehumbert for fruitful
discussions early in the development of this project. We acknowledge
input from Francis MacDonald on Snowball geochronology, John Higgins
on weathering parameterizations, and Bob Wordsworth on the effect of
CO$_2$ clouds on the habitable zone. Jacob Bean, Albert Colman, Itay
Halevy, Guillaume Le Hir, Edwin Kite, Daniel Koll, Arieh Konigl, and
an anonymous reviewer provided helpful comments on early drafts of
this paper. We thank Bj\"orn Benneke for sharing an early draft of his
manuscript.

%


\end{document}